\documentclass[reprint,showpacs,preprintnumbers,amsmath,amssymb,prc,floatfix]{revtex4-1}
\usepackage{float}
\usepackage{color}
\usepackage{graphicx}
\usepackage{dcolumn}
\usepackage{bm}
\usepackage{xcolor}
\usepackage{multirow}
\usepackage{adjustbox}
\usepackage{comment}
\usepackage[colorlinks,citecolor=blue,linkcolor=red,anchorcolor=blue,filecolor=blue,urlcolor=blue]{hyperref}
\begin{document}
\title{Relativistic mean-field study of alpha decay in superheavy isotopes with 100 \texorpdfstring{$\leq$ Z $\leq$}-120}

\author{Nishu Jain$^1$}
\email{nishujain1003@gmail.com}
\author{M. Bhuyan$^{2}$}
\email{bunuphy@um.edu.my}
\author{Raj Kumar$^1$}
\email{rajkumar@thapar.edu}
\affiliation{$^1$School of Physics and Materials Science, Thapar Institute of Engineering and Technology, Patiala, Punjab 147004, India}
\affiliation{$^2$Center for Theoretical and Computational Physics, Department of Physics, Faculty of Science, Universiti Malaya, Kuala Lumpur 50603, Malaysia}

\date{\today}
\begin{abstract}
The $\alpha$-decay half-lives of superheavy nuclei with $100 \leq Z \leq 120$ are comprehensively analyzed using the axially deformed relativistic mean field (RMF) formalism for the NL3$^*$ parameter set. We employ RMF binding energies to determine the $\alpha$-decay energies and make a comparison with both the available experimental data and the theoretical results obtained from the global nuclear mass model WS4. The four distinct formulae, specifically the modified scaling law Brown, modified Viola-Seaborg, Yibin {\it et al.} formula, and its modified form are used to calculate the decay half-lives and examine the numerical correlation between the half-life ($T_{1/2}$) for each $\alpha$-decay energy. We notice that $T_{1/2}$ is significantly dependent on the decay formula in terms of isospin asymmetry and decay energy. We also noticed that modified scaling law Brown formula estimates of half-lives agreed comparatively better with the experiment as compared to others. Moreover, the present investigation provides significant information on the stability of the superheavy island considered for ongoing and/or future experiments. 
\par 
\end{abstract}
\maketitle
\vspace{-0.5cm}
\section{Introduction}\label{sec1}
Recent advancements in superheavy nuclei (SHN) research have generated significant interest, establishing it as a fascinating field within modern nuclear physics \cite{sobi16,manj19,ogan17}. Due to their intricate synthesis and numerous unexplored properties, these nuclei pose an exciting challenge for experimental studies. Various attempts to synthesize SHN have been carried out by Oganessian {\it et al.} \cite{ogan09,ogan12} and the Refs. therein. Additionally, the stability of SHN, primarily influenced by shell effects, has long been a critical concern in nuclear physics \cite{mose69,bend01,cwio96}. Alpha decay has been pivotal in the advancement of nuclear physics and its practical applications since the discovery of radioactive decay in the early $20^{th}$ century. This is because alpha decay is connected to various fundamental aspects of nuclear physics and astrophysics. The demand for studying alpha decay has increased due to rapid advancements in modern detector technology, the production of superheavy nuclei, and the use of radioactive beams. By comprehending the process of alpha disintegration, researchers gain insight into various characteristics of nuclear structure, including deformation, spin-parity, energy levels, ground state, and the coexistence of nuclear shapes \cite{seif11,carr14,poen06,park05}. Additionally, studies on decay processes serve as essential tools for identifying newly synthesized superheavy nuclei \cite{mori07,ogan11}. Concepts such as proton radioactivity and cluster radioactivity also contribute to understanding the nuclear structure, and all these phenomena can be represented through a single framework of barrier penetration \cite{dong09,poen18,deng19}.

Recent experimental studies have focused on exploring $\alpha$-emitters with remarkably long half-lives, particularly within the artificially synthesized superheavy nuclei (SHN) that occupy the upper right corner of the nuclear landscape. Multiple research facilities worldwide, including JINR-FLNR (Dubna), RIKEN (Japan), and GSI (Darmstadt), have successfully synthesized nuclei with atomic numbers Z = 118 \cite{nave20}. The production of superheavy elements involves two primary fusion-evaporation processes: cold and hot fusion reactions, which are extensively studied and employed in laboratories around the world \cite{hoff00,ogan07a}. Extending the boundaries of the nuclear chart's upper limit is a fundamental objective that motivates modern nuclear physics research. Scientists in various laboratories are investigating the synthesis of elements beyond element $Og$, employing different combinations of projectiles and targets \cite{ogan09,khuy20} to explore the unknown regions of the nuclear landscape. Theoretical models, including the generalized liquid drop model (GLDM) and the two-potential approach (TPA), are used to study $\alpha$-decay mechanisms \cite{xu06,bhat08,dong10,deli06,pelt07} and various empirical methods to determine decay half-lives \cite{qian12,akra18}. The ongoing search for naturally occurring superheavy elements with long lifetimes has driven scientific attempts in nuclear, atomic physics, and chemistry. The ongoing quest for naturally occurring superheavy elements with prolonged lifetimes has led researchers to focus on the upper right-hand corner of the nuclear chart, where SHN with $Z \geq 104$ \cite{hoff00,seab90}. The initial theoretical results indicated the existence of superheavy magic numbers at Z = 114 and N = 184 \cite{myer66,sobi66,viol66}. Theoretical advancements have predicted proton magic numbers of 114, 120, 124, or 126 and neutron magic numbers of 172 or 184 \cite{gupt97,patr99,ahma12,tani20}. Unlike stable systems, superheavy nuclei tend to exhibit extended half-lives due to the influence of single-particle levels and strong Coulomb repulsion, resulting in a diffuse shell structure \cite{bend01,jerk18,dull18}. The novel isotopes are typically identified from theoretical predictions and evidence from experiments by examining their distinctive $\alpha$-decay chains.

Even with the challenges it presents, a thorough understanding of the concepts behind the ``Island of stability" and the doubly magic spherical nucleus can be achieved by studying the structural characteristics of nuclei within the superheavy mass region. Theoretical models often estimate the $\alpha$-decay half-lives of nuclei for which experimental data is lacking. The present investigation is a continuation of our previous research \cite{bisw21,jain22}, where our objective is to employ a microscopic model to determine the decay characteristics of the superheavy island. This extensive study examines the $Q$-values and decay half-lives of even-even nuclei with atomic numbers ranging from ($100 \leq Z \leq 120$). In this specific study, the decay half-life of the chosen superheavy nuclei is determined using four different semi-empirical formulas: the modified Viola-Seaborg formula, the modified scaling law Brown formula, the modified Yibin {\it et al.} formula, and the Yibin {\it et al.} formula \cite{akra19}. The relativistic mean field with NL3$^*$ parameter \cite{lala09}, which is the refitted version of NL3 force parameter \cite{lala97}, is employed for the present investigations. Furthermore, it is essential to investigate the numerical dependence of the half-life for each $\alpha$-decay energy. The results are then compared with those from the Finite Drop Liquid Model (FRDM), the Global Nuclear Mass Model (WS4), and experimental data, wherever available. The $T_{1/2}$ values in the superheavy mass region are commonly calculated using the abovementioned formulas. It is assumed that these decay formulae significantly influence the half-lives of all selected SHN (Superheavy Nuclei). Therefore, we are particularly interested in examining the results produced by these empirical formulas regarding the $Q$-value for $\alpha$-decay. Furthermore, we computed the systems' probable mean or standard deviation with available experimental data for the superheavy island. This allows us to assess the applicability and accuracy of the model. Ultimately, our objective is to identify the most stable and reliable Superheavy Nuclei (SHN) through this research. By doing so, we hope to contribute to the understanding and identification of nuclei in the superheavy mass region.\\ \\
The following is an outline for this paper: A summary of the RMF theory is provided in Section \ref{theory}. The discussion of results for nuclear structure and its decay characteristics is mentioned in Section \ref{result}. In Section \ref{summary}, a summary of the results of the current work is presented.
\section{Theoretical formalism}
\label{theory} \noindent
The nuclear structure and infinite characteristics of nuclear matter have long been described using the mean-field QHD treatment \cite{bogu77,sero86,bhuy11,bhuy15,type99,niks02}. Relativistic mean-field theory proposes that the nucleus is a complex interaction of nucleons (\textit{N} and \textit{Z}) that communicate by exchanging mesons and photons \cite{sero86,ring96,paar07,bend03}. Characterizing meson field effects as point-like interactions between constituent elements is essential for accurately representing the saturation properties of infinite nuclear matter \cite{type99,niks02,niks08,fuch95}. The non-linear coupling factors or density-dependent coupling constants are also included \cite{bogu77,broc92,niko92,burv02}. The mathematical formulation of a many-body system within the framework of relativity is expressed through the relativistic Lagrangian
\cite{bogu77,sero86,bhuy15,ring96,paar07,rein89,vret05,bhuy18,patr09,niks11}:
\begin{eqnarray}
{\mathcal L}&=&\overline{\psi}\{i\gamma^{\mu}\partial_{\mu}-M\}\psi +{\frac12}\partial^{\mu}\sigma
\partial_{\mu}\sigma -{\frac12}m_{\sigma}^{2}\sigma^{2} \nonumber \\
&-&{\frac13}g_{2}\sigma^{3} -{\frac14}g_{3}\sigma^{4} -g_{s}\overline{\psi}\psi\sigma 
-{\frac14}\Omega^{\mu\nu}\Omega_{\mu\nu} \nonumber \\
&+&{\frac12}m_{w}^{2}\omega^{\mu}\omega_{\mu}
-g_{w}\overline\psi\gamma^{\mu}\psi\omega_{\mu} 
-{\frac14}\vec{B}^{\mu\nu}.\vec{B}_{\mu\nu} \nonumber \\ &+&\frac{1}{2}m_{\rho}^2 \vec{\rho}^{\mu}.\vec{\rho}_{\mu} 
-g_{\rho}\overline{\psi}\gamma^{\mu}
\vec{\tau}\psi\cdot\vec{\rho}^{\mu}-{\frac14}F^{\mu\nu}F_{\mu\nu} \nonumber \\
&-&e\overline{\psi} \gamma^{\mu}
\frac{\left(1-\tau_{3}\right)}{2}\psi A_{\mu}.
\label{lag}
\end{eqnarray}
here, $\psi$ is the Dirac spinors of the nucleons. The coupling constants for the $\sigma$, $\omega$ and $\rho$ mesons are represnted by $g_\sigma(m_\sigma)$, $g_\omega(m_\omega)$, and $g_\rho(m_\rho)$\, respectively.  The third component of isospin and isospin is symbolized as $\tau_3$, and $\tau$ respectively. The self-interacting non-linear $\sigma$-meson field in relativistic mean-field theory is governed by non-linear coupling constants $g_2$ and $g_3$, while the photon field is characterized by the coupling constant $e^2/4\pi$. The mass of the nucleon is represented by $M$, while the electromagnetic field is denoted as $A$. The vector field tensors are given by
\begin{eqnarray}
F^{\mu\nu} = \partial_{\mu} A_{\nu} - \partial_{\nu} A_{\mu} \nonumber \\
\vec{B}^{\mu\nu} = \partial_{\mu} \vec{\rho}_{\nu} - \partial_{\nu} \vec{\rho}_{\mu}\nonumber \\
\Omega_{\mu\nu} = \partial_{\mu} \omega_{\nu} - \partial_{\nu} \omega_{\mu}.
\end{eqnarray}
for the photon fields, $\vec{\rho}_{\mu}$, and $\omega^\mu$ respectively. The aforementioned Lagrangian density is used to determine the field equations for mesons $\&$ nucleons. The advancement of lower and upper Dirac spinor components, along with the boson fields, can be attributed to an initial deformation of $\beta_0$ within an axially deformed harmonic oscillator basis. The numerical solution of the set of coupled equations is achieved by utilizing a self-consistent iteration method. The traditional harmonic oscillator formula can be used to determine the centre-of-mass motion energy \cite{long04}: $E_{c.m.}=\frac{3}{4}(41A^{-1/3})$. Here $A$ is the mass number. To determine the overall binding energy and other pertinent observables, the standard relationships described in Ref. \cite{pann86} are utilized. The RMF framework is utilized in conjunction with a constant gap BCS method (with NL3$^*$ parameter set) for this study \cite{lala97,patr09,niks11,kara10}.
\begin{figure*}[ht]
\begin{center}
\includegraphics[width=2.0\columnwidth]{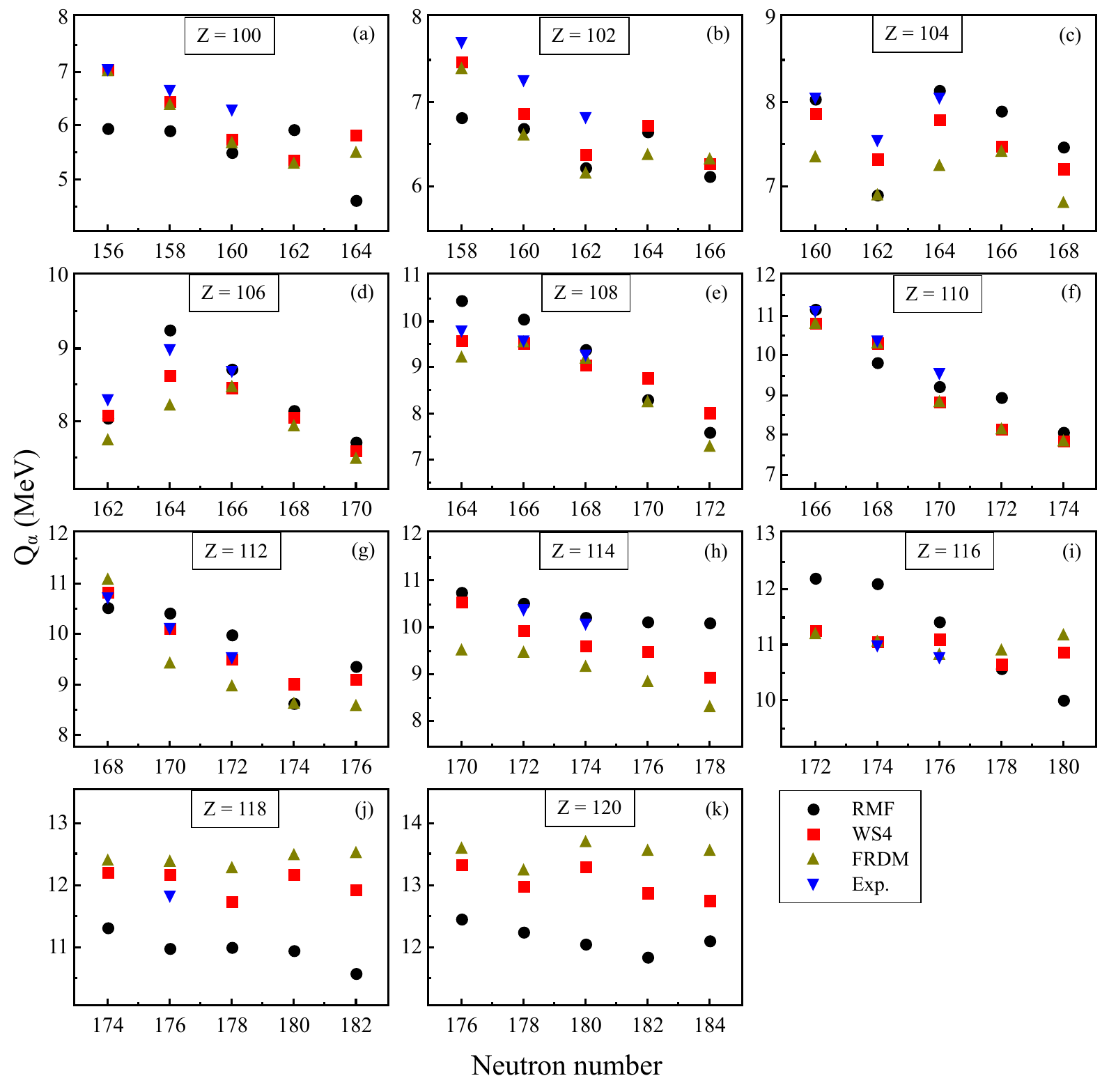}
\caption{\label{fig1} (Color online) The $\alpha$-decay energy for the isotopic chain of Z = 100 to 120 is calculated via RMF (NL3$^*$) and compared to the FRDM results, \cite{moll19} WS4 predictions, \cite{wang14} and the existing experimental data.\cite{wang12}}
\end{center}
\end{figure*}
\begin{figure*}[ht]
\begin{center}
\includegraphics[width=2.0\columnwidth]{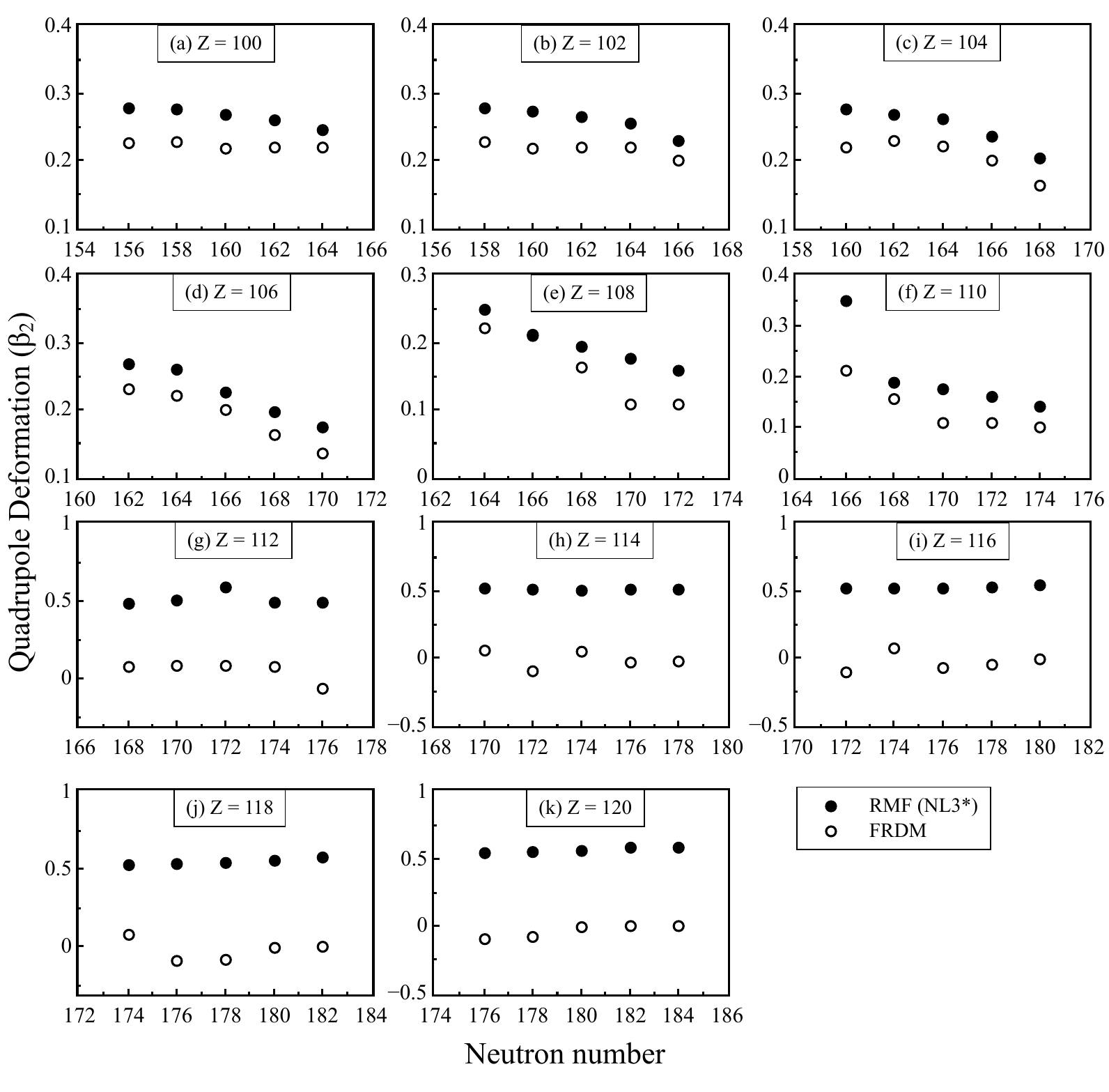}
\caption{\label{fig6} (Color online) The quadrupole deformation parameter ($\beta_2$) for the isotopic chain of Z = 100 to 120 nuclei by using relativistic mean-field approach with NL3$^*$ are given along with the FRDM predictions \cite{frdm97,frdm16}.}
\end{center}
\end{figure*}
\begin{figure*}[ht]
\begin{center}
\includegraphics[width=2.0\columnwidth]{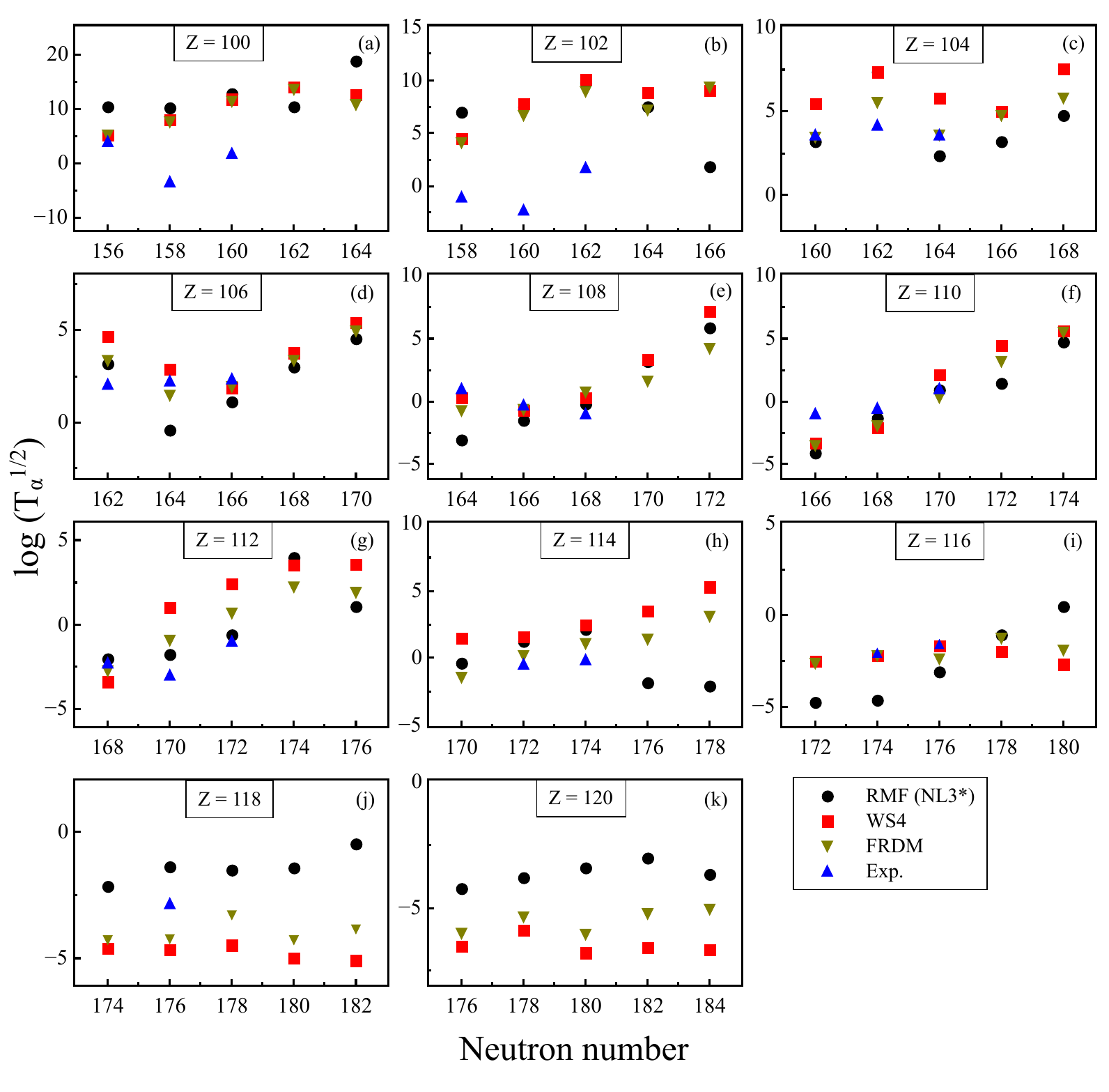}
\vspace{-0.6cm}
\caption{\label{fig2} (Color online) The estimated $T_{1/2}^{\alpha}$ for the isotopic chain of Z = 100 to 120 nuclei using modified Viola-Seaborg formula (MVS).\cite{akra19} The results are further compared with the other theoretical and experimental results as well.}
\end{center}
\end{figure*}
\begin{figure*}[ht]
\begin{center}
\includegraphics[width=2.0\columnwidth]{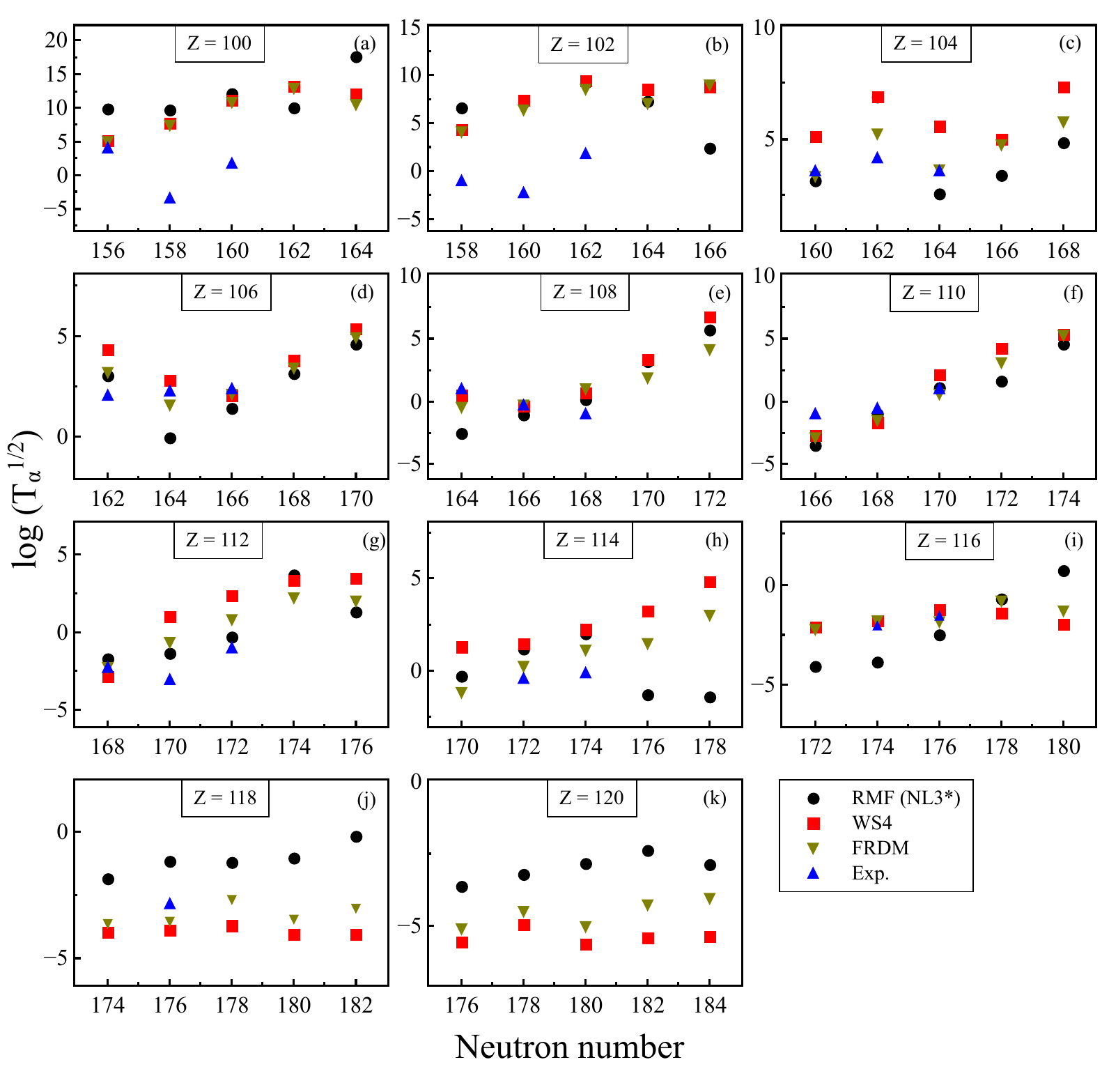}
\vspace{-0.6cm}
\caption{\label{fig3} (Color online) Same as in Fig. \ref{fig2} but for modified scaling law Brown formula (MSLB) \cite{audi17}.}
\end{center}
\end{figure*}
\begin{figure*}[ht]
\begin{center}
\includegraphics[width=2.0\columnwidth]{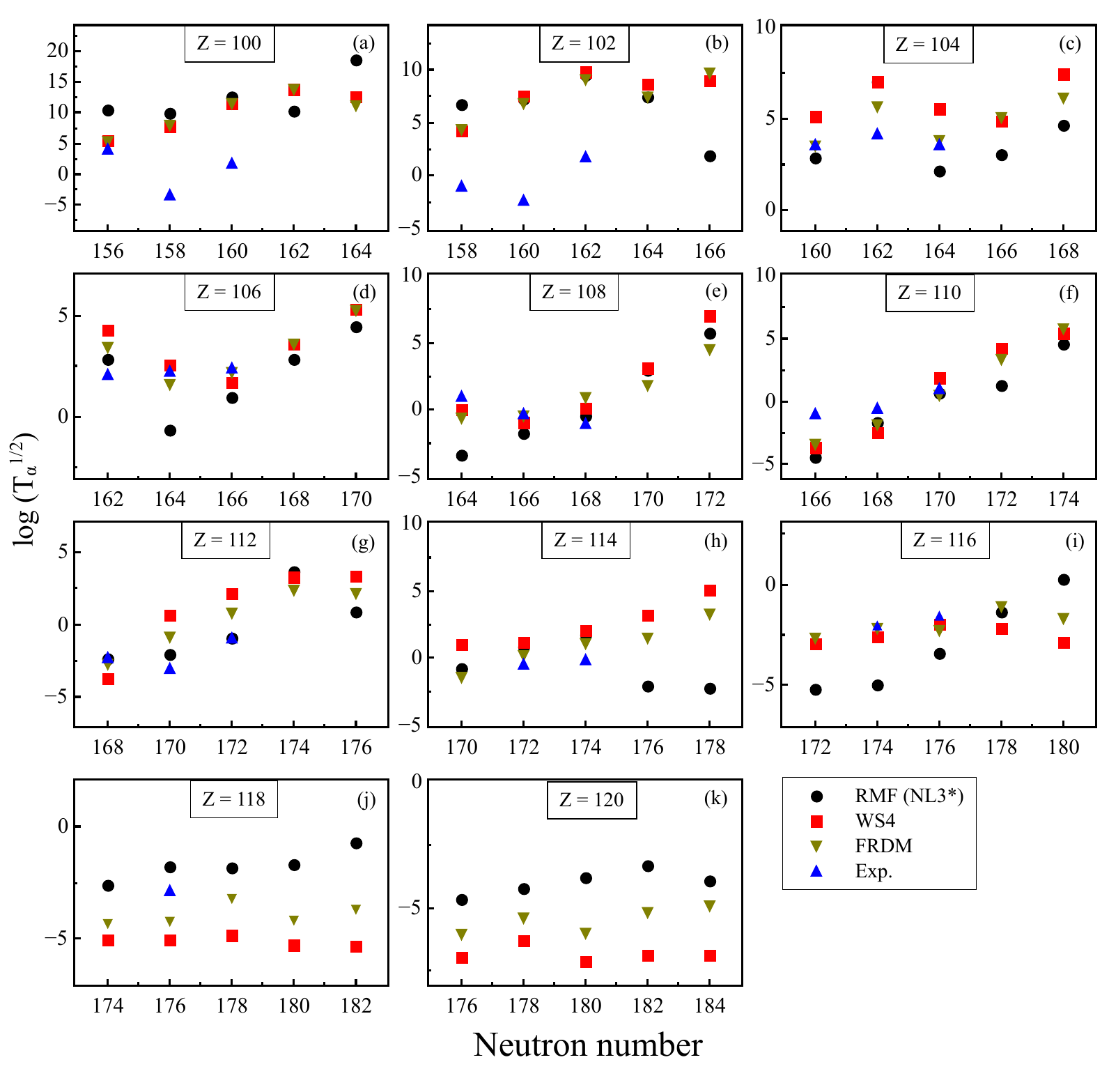}
\vspace{-0.6cm}
\caption{\label{fig4} (Color online) Same as in Fig. \ref{fig2} but for Yibin {\it et al.} formula (YQZR) \cite{audi17}.}
\end{center}
\end{figure*}
\begin{figure*}[ht]
\begin{center}
\includegraphics[width=2.0\columnwidth]{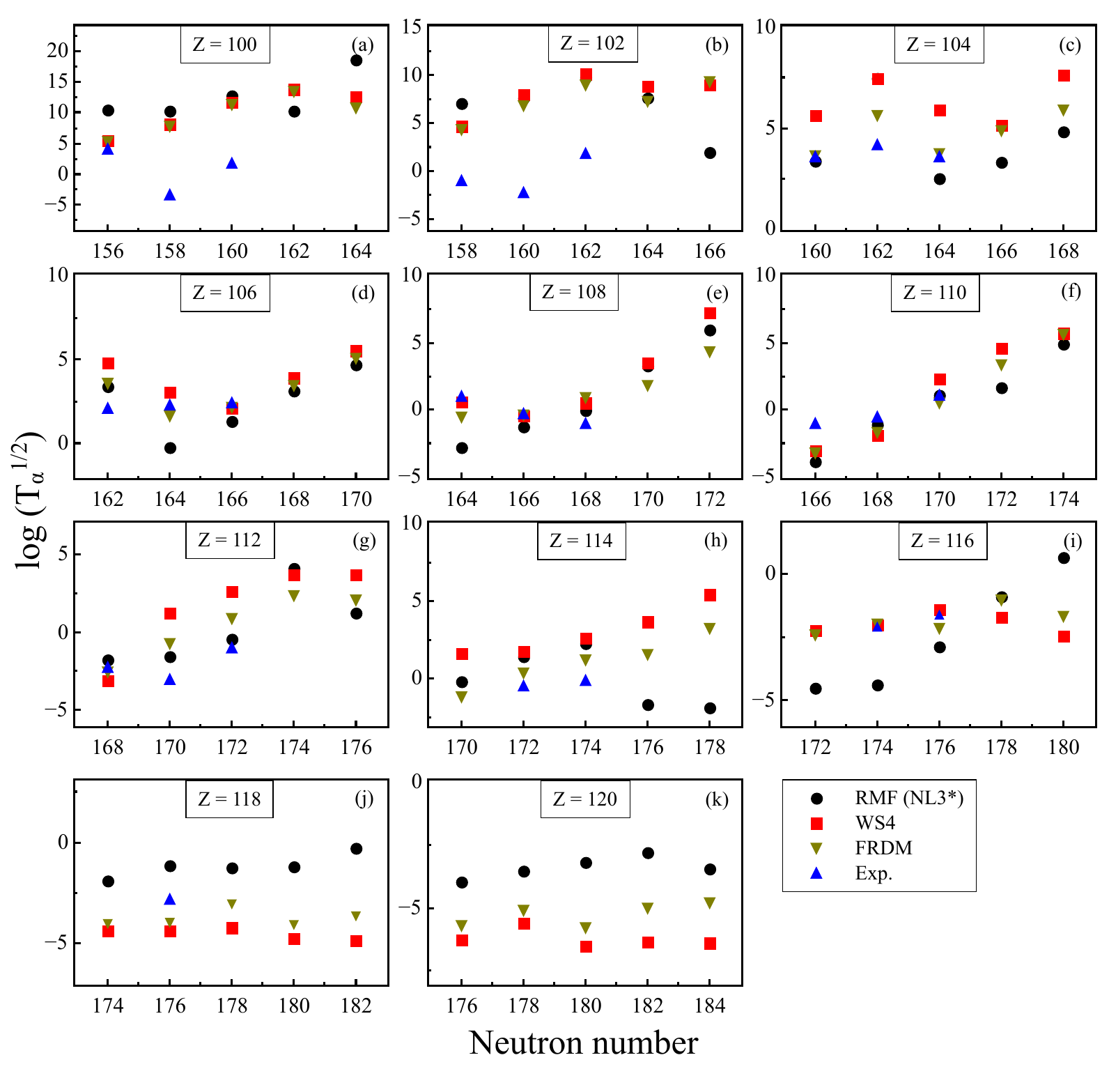}
\vspace{-0.6cm}
\caption{\label{fig5} (Color online) Same as Fig. \ref{fig2} but for modified Yibin {\it et al.} formula (MYQZR) \cite{audi17}.}
\end{center}
\end{figure*}

\section{Results and Discussions} 
\label{result}
Accurate prediction of nuclear masses across the nuclear chart is paramount for physicists. Nuclear binding energies \cite{lunn03} play a pivotal role in identifying magic numbers and regions of stability. A self-consistent approach is necessary to achieve a convergent solution for ground and excited states in the superheavy region, involving the variation of the initial deformation $\beta_0$ \cite{lala09,ring96,lala99}. In this context, merging fermionic and bosonic ground-state solutions requires $N_F$ = $N_B$ = 20 major shells. The numerical integration involves 20 mesh points in Gauss-Laguerre and 24 mesh points in Gauss-Hermite. This study employs the well-established and recently developed non-linear NL3$^*$ parameter sets \cite{lala97}, which are particularly suitable for describing exotic features of the nuclear landscape, including the drip line and superheavy regions. An intriguing possibility for further investigation is exploring the potential of the relativistic mean field approach using the NL3$^*$ parameter set and assessing its accuracy in predicting the bulk properties of superheavy nuclei throughout the nuclear chart. This study elucidates how the force parameter influences the structural properties of superheavy nuclei. Specifically, the investigation focuses on nuclear decay for even-even superheavy nuclei with atomic numbers $Z$ ranging from 100 to 120 and neutron numbers between $156 \leq N \leq 184$. Subsequent sections will examine the alpha decay energy ($Q_\alpha$)-values and half-life ($T_{1/2}$) associated with the decay of these superheavy nuclei.
\subsection{Decay Properties of Superheavy Nuclei}
\noindent
The $\alpha$-emission is one of the principal decay channels observed in the synthesized superheavy nuclei, making it a fundamental process for understanding their nuclear structure and stability. The $Q_\alpha$ values and corresponding half-lives are the main sources of information about the nuclear structure of heavy nuclei. By studying $\alpha$-emission and its characteristics, researchers can gain valuable insights into the nuclear structure, behaviour, and stability of heavy nuclei, providing essential information for nuclear physics and astrophysics. Previous studies have successfully investigated the ground state characteristics of actinides and superheavy nuclei, employing various relativistic parameterizations \cite{patr99,ahma12,tani20,bisw21,patr09,zhan06,sahu11,bhuy15a,bhuy18a}. These investigations have shown that the NL3$^*$ parameter set exhibits reasonable agreement with experimental data \cite{bisw21,jain22}. However, due to the unavailability of experimental data for Z $>$ 118 nuclei, we will employ relativistic mean-field calculations with the NL3$^*$ parameter set to determine the decay energy of these nuclei. To determine the decay half-lives of the selected superheavy nuclei, we have used four semi-empirical formulas: the modified scaling law Brown formula (MSLB), modified Viola-Seaborg formula (MVS), Yibin et al. formula (YQZR), and the modified Yibin et al. formula (MYQZR) \cite{akra19}. Our previous analysis demonstrated that the NL3$^*$ parameter set yields highly satisfactory results compared to experimental data for identified superheavy nuclei \cite{bisw21,jain22}. Thus, applying these four semi-empirical formulas to compute decay half-lives will be fascinating and critical as we proceed into unexplored areas of nuclear research. \\
\noindent
\textbf{The $\alpha$-decay Energy ($Q_\alpha$)}:
We comprehensively analyze the $\alpha$-decay properties of 55 SHN, considering various $\alpha$-decay chains. The $Q_\alpha$ and $T_{1/2}$ estimates for these nuclei are examined and gain insights into their nuclear stability. This decay energy is crucial because it determines the energy barrier that the parent nucleus must overcome to transform into the daughter nucleus through $\alpha$-emission. The value of $Q_\alpha$ affects the decay probability and, subsequently, the half-life of the nucleus, which is a key indicator of its stability. The stable nuclei and those exhibiting shell/sub-shell closure are often identified by their decay half-lives compared to their neighbouring nuclei. For example, a nucleus possesses a half-life longer than the typical decay timescale and the shell closures are comprehensive with certain magic numbers of protons and neutrons suggesting higher stability. The decay energy $Q_\alpha$ is calculated by using the relation, $Q_\alpha (N, Z) = BE (2, 2) + BE (N-2, Z-2) - BE (N, Z)$. Here, BE $(2, 2)$, BE $(N, Z)$, and BE $(N-2, Z-2)$  are the binding energies of $\alpha$-particle (BE = 28.296 MeV), parent, and daughter nuclei, respectively. With these estimated binding energies and applying the $Q_\alpha$ equation, we can determine the decay energy for each $\alpha$-decay process. 

The $Q_\alpha$ values are calculated by employing the binding energies from RMF (NL3$^*$) for $^{256,258,260}$Fm, $^{260,262,264}$No, $^{264,266,268}$Rf, $^{268,270,272}$Sg, $^{272,274,276}$Hs, $^{276,278,280}$Ds, $^{280,282,284}$Cn, $^{286,288}$Fl, $^{290,292}$Lv, $^{294}$Og nuclei for which experimental data is available. We further extend this study to several SHEs such as $^{262,264}$Fm, $^{266,268}$No, $^{270,272}$Rf, $^{268,270,272}$Sg, $^{278,280}$Hs, $^{282,284}$Ds, $^{286,288}$Cn, $^{284,290,292}$Fl, $^{288,294,296}$Lv, $^{292,296,298,300}$Og, and $^{296-304}$120 for which experimental data is not available. 
In Fig. \ref{fig1}, the alpha decay energy $Q_\alpha$-values from RMF (NL3$^*$) are compared with predictions from the Finite Range Droplet Model (FRDM) \cite{moll19}, WS4 results \cite{wang14}, and experimental data \cite{wang12}. Notably, for massive isotopes, the $Q_\alpha$ values obtained from FRDM and WS4 exhibit relatively higher values when compared to the results obtained from RMF calculations. This discrepancy highlights the importance of considering the shape transition exhibited in the ground state configuration of these heavy nuclei, which is not adequately accounted for in the FRDM models, especially for high mass numbers. 

The quadrupole deformation parameters are calculated for the RMF(NL3$^*$) parameter set and are shown with the FRDM predictions \cite{frdm97,frdm16} shown in Fig. \ref{fig6}. It is worth noting that the quadrupole deformation parameters are a crucial determinant of nuclear shapes in both the ground and intrinsic excited states. From Fig. \ref{fig6}, one can notice that the RMF (NL3$ ^*$) results for some mass regions did not agree with the FRDM predictions. For instance, considering the isotopes within the range $Z$ = 100–110, the calculated $\beta_2$-value decreases as the mass number increases in a range of 0.1 $\le \beta_2 \le$ 0.3, which is in agreement with the FRDM predictions. For the $Z$ = 112-120 range, there is a significant rise in the $\beta_2$ value, indicating a transition from a slightly deformed prolate configuration to a highly or superdeformed prolate shape. This transition is explicitly observed for the NL3$^*$ parameter set. On the other hand, FRDM only exhibits a slight oblate configuration in a few isotopes along with the same pattern, showing no change in shape at all. It will be interesting to observe the low-lying intrinsic excited state of the spherical configuration followed by a superdeformed and/or hyperdeformed ground state, even though no experimental data is available. Following the Refs. \cite{frdm16,rama01,bisw21,patr09,ahma12} suggest that a more comprehensive understanding of the ground state structure could be attained by exploring additional degrees of freedom, specifically octupole and hexadecapole deformations \cite{lu16}. This multifaceted approach enhances the depth of our exploration, offering insights into the intricate interplay of various deformations that dictate the shape of nuclei in both ground and excited states. 

To assess the accuracy of our calculations, we have computed the mean deviation (MD) in decay energies for mass estimates using RMF (NL3$^*$), WS4, and FRDM, and found the corresponding values to be 0.189, 0.164, and 0.387, respectively. Additionally, mean deviations for RMF (NL3$^*$), when compared with FRDM and WS4 using additional theoretical models, are found to be 0.033 and 0.097, respectively. These results suggest that all computed mean deviations fall within an acceptable range, indicating the reliability of the theoretical frameworks used in the present analysis. In a more detailed analysis, the mean deviation for FRDM mass is more as compared to RMF (NL3$^*$), though WS4 exhibits a slightly smaller value. This implies that the decay energy obtained from RMF (NL3$^*$) calculations may be slightly underestimated compared to experimental data. Meanwhile, when using other theoretical models, the resulting decay energies show slightly higher values concerning the experimental results. However, despite these small deviations, the relative mean deviations of RMF with the NL3$^*$ parameter and other theoretical frameworks from experimental data suggest a reasonable agreement, thereby reinforcing the validity of our approach. Utilizing the computed decay energies from RMF (NL3$^*$) and other theoretical frameworks, we can deduce the decay half-life of the unknown region in the superheavy island, providing valuable insights into this unexplored region.\\ 

\noindent
\textbf{The $\alpha$-decay half-life ($T_{1/2}^\alpha$)}:
Measuring $\alpha$-decay energy, decay widths and corresponding half-life has been ongoing since the early 1900s, with Geiger and Nuttall making a significant discovery in 1911. The recent study of Ref. \cite{geig11} has identified a direct relationship between the $log T_{1/2}^\alpha$ and $Q_{\alpha}^{-1/2}$, which paved the way for further investigations to explore and refine the underlying principles governing $\alpha$-decay. Building upon the initial breakthrough, subsequent researchers have employed different approaches to improve the accuracy and reliability of estimating $\alpha$-decay half-lives. These approaches include analytical methods \cite{Roye00,Roye10}, semi-empirical techniques \cite{Poen07,Manj17}, and empirical models \cite{Akra17,Akra18b}. Each approach aimed to modify and enhance the existing formula to better correlate experimental observations with estimated $\alpha$-decay half-lives. In the present study, we have adopted a set of four recently formulated semi-empirical equations to estimate the alpha decay half-lives of specific superheavy nuclei (SHN) under investigation. \\
{\bf Modified Viola-Seaborg formula (MVS)}: The MVS formula is expressed as:
\begin{eqnarray*}
\log_{10} T_{1/2}^{MVS} (s) = (aZ+b)Q^{-1/2}_\alpha+cZ+d+eI+fI^2 . 
\end{eqnarray*}
The nuclear asymmetry term $I = (N-Z)/A$, describes the difference between the number of neutrons ($N$) and protons ($Z$) in a given nucleus of mass number ($A$). Here, $I$ $\&$ $I^2$ which are two asymmetry-dependent terms that are linearly associated to the logarithm of $T_{1/2}^\alpha$ were also added to the Viola-Seaborg formula. The fitting parameters for even-even nuclei are taken from Ref. \cite{akra19} and are given as $a$, $b$, $c$, $d$, $e$ and $f$ possesses 1.53, 5.69, -0.17, -36.54, 6.08, and -39.58 values, respectively.  \\
{\bf Modified scaling law of Brown (MSLB)}: The modification in the scaling law of the Brown formula has been constructed by adding the asymmetry-dependent ($I$ and $I^2$) terms which are linearly related to the logarithm of $\alpha$-decay half-lives, and the Modified scaling law of the Brown formula is given as,
\begin{eqnarray*}
\log_{10} T_{1/2}^{MSLB} = aZ_d^{0.6}Q_{\alpha}^{-1/2} + b + cI + dI^2.
\end{eqnarray*}
Here, Z$_d$ is the proton number of the daughter nucleus and the term `asymmetry' ($I$) refers to the difference between $N$ and $Z$ in a nucleus. The fitting parameters for all even-even nuclei are given as $a, b, c,$ and $d$ are  9.04, -49.63, 6.88, and -4.21, respectively \cite{akra19}.\\ 
{\bf YQZR formula (YQZR)}: A new empirical formula, proposed by Yibin {\it et al.}, builds upon the formula developed by Ni {\it et al.} (NRDX) by introducing a new parameter so called angular momentum $L$ to enhance its predictive power. This modified formula, known as the YQZR formula, can be expressed as follows:
\begin{eqnarray*}
\log_{10} T_{1/2}^{YQZR} = a\sqrt{\mu} Z_1Z_2 Q^{-1/2} + b\sqrt{\mu} (Z_1Z_2)^{1/2}  \nonumber\\
+ c \frac{L(L+1)}{\mu \sqrt{Z_1Z_2}A^{1/6}} + d. 
\end{eqnarray*}
In YQZR formula, $Z_1$, $Z_2$, and $\mu$ represent the atomic numbers of daughter nuclei, alpha particles, and reduced mass of the compound system. Additionally, the formula includes four sets of free parameter coefficients ($a$, $b$, $c$ and $d$) taken from Ref. \cite{akra19} having values 0.40, -1.50, 0.00, and -11.71, respectively for $even-even$ nuclei.\\ 
{\bf Modified YQZR formula (MYQZR)}: The YQZR formula was further modified to include two additional terms, $I$ and $I^2$, which are linearly related to the logarithm of alpha-decay half-lives. These modifications enhance the formula's predictive power and provide a more accurate estimate of $T_{1/2}$ and the MYQZR is:
\begin{eqnarray*}
\log_{10} T_{1/2}^{MYQZR} = a\sqrt{\mu} Z_1Z_2 Q^{-1/2} + b\sqrt{\mu} (Z_1Z_2)^{1/2} \nonumber\\
+ c \frac{L(L+1)}{\mu \sqrt{Z_1Z_2}A^{1/6}} + d + eI + fI^2
\label{vss-decay}
\end{eqnarray*}
The modified form contains six fitting parameters (including four constants from YQZR formula refitted) $a$, $b$, $c$, $d$, $e$ and $f$ for $even-even$ having values 0.40, -1.45, 0.00, -14.87, 13.39, and -61.47, respectively \cite{akra19}.  

The four semi-empirical formulae considered in the present calculation are relatively competent in determining the alpha-decay half-lives ($T_{1/2}$) of the superheavy island. This study investigated the half-lives of even-even isotopes within the range of atomic numbers $Z$ = 100 to 120. The data obtained from this study are presented in Figs. \ref{fig2} to \ref{fig5}. To accomplish this, predictions from three distinct models, namely RMF (Relativistic Mean Field), FRDM (Finite-Range Droplet Model), and WS4 (Weizs$\ddot{a}$cker-Skyrme), are employed to obtain the $Q$-values for these nuclei. The $Q$-values represent the energy released during the alpha-decay process and play a pivotal role in determining the alpha-decay half-lives. The variation in these $Q$-values is minimal, resulting in only a small variation in the computed $\log_{10} T_{1/2}$ values when used as inputs in the analytical half-life formulas. While the decay energy and the isospin asymmetry have been truly identified as notable parameters influencing $\alpha$-decays, it is worth noting that the complexities around the superheavy region are yet to be fully understood especially at higher Z numbers \cite{hofm15,giul19}. These unknown complexities limit the construction and accurate extrapolation of the available formulae. Thus, the knowledge of most of the participating degrees of freedom in this region still awaits sufficient theoretical and experimental evidence. We have performed similar calculations in our recent work \cite{thee23} for the predictions of decay half-lives using different semi-empirical formulae. In the present work, we initially focused on evaluating the $T_{1/2}$ values for already-known superheavy elements, for which experimental data \cite{audi17} were available and subsequently, the study extended to the unexplored superheavy region. 

The study conducted a detailed analysis of the $T_{1/2}$ for a specific $Q_\alpha$-values of 12.10 MeV for $^{304}120$. This analysis revealed a significant discrepancy in the half-life calculations using the formulae. For the $^{304}120$ nuclei with the given $Q$-value, the four formulae (MVS, MSLB, YQZR, and MYQZR) produced four distinct values of $\log_{10} T_{1/2}$, like -3.66, -2.87, -3.93 and -3.45 seconds, respectively. Similar predictions are also drawn for all considered nuclei. A comparison of our estimates with those obtained from the FRDM \cite{moll19} and WS4 \cite{wang14} highlights the substantial differences in $\log_{10} T_{1/2}$ values, particularly for larger mass nuclei. However, it is worth noting that the values from RMF, FRDM, and WS4 models fall within an acceptable range, especially for nuclei with relatively lower mass numbers. The accuracy of the $\log_{10} T_{1/2}$ values largely depends on the effectiveness of the alpha-decay formula used in the calculations. Since all the employed formulae were modified by incorporating new coefficients \cite{akra19} based on the experimental $\alpha$-decay half-life data of 356 nuclei. The accuracy of these formulae was determined by comparing them with the experimental results. 

These formulae appear to be well-suited for the mass region within the range of 105 $\leq$ Z $\leq$ 120. However, it exhibits limitations in accurately reproducing results in the lower mass region. Interestingly, as it extends into the specified range, it demonstrates its capability to reproduce results effectively. This suggests that the formula's applicability may be context-dependent and may require adjustments or considerations when applied to different mass regions. To assess the reliability of our calculations, we compared our results with previous results \cite{jain22}. This analysis revealed a remarkable correlation between the experimental data and the results obtained from the proposed semi-empirical formulas. To further validate the consistency of these formulas, we computed the standard deviation relative to other existing theoretical outcomes. The present study demonstrated that the MSLB results give a better match with the experimental data, exhibiting a standard deviation of 4.15, while other semi-empirical formulas had slightly lower values than the experimental data. The results obtained from MSLB can be employed to predict the acceptable $\alpha$-decay half-lives for the investigated superheavy nuclei. Interestingly, the analysis also identified new isotopes with specific shell/sub-shell closures, leading to extended $\log_{10} T_{1/2}$ values. This finding is particularly significant as it supports the validity of assumptions made in earlier research studies \cite{ahma12,tani20,bisw21,jain22,zhan06}. Notably, isotopes with neutron numbers $N$ = 162, 164, and 174-184 exhibited longer half-lives than other superheavy isotopes. This expectation of longer half-lives for certain isotopes is paramount in the context of upcoming experimental synthesis efforts. 

\section{Summary and Conclusions}\label{summary}
We employed the RMF formalism with the NL3$^*$ parameter set to explore the $\alpha$-decay properties of even-even nuclei ranging from $Z$ = 100 to 120 in this research. Our findings demonstrate that the calculated decay energy and $\log_{10} T_{1/2}$ are in good accordance with experimental data, suggesting that the NL3$^*$ parameterization is a suitable tool for simulating an uncharted superheavy region of the nuclear landscape. Our method involved calculating the $\alpha$ decay energies utilizing $Q$-values obtained from the binding energies of parent and daughter nuclei with the RMF (NL3$^*$) model and determining the $\alpha$ decay half-lives using four different semi-empirical formulas. We compared the outcomes with experimental data, as well as with the macroscopic-microscopic FRDM and the WS4 mass model. Furthermore, we determined the mean deviations for the RMF (NL3$^*$) results using various theoretical predictions and experimental data, demonstrating excellent agreement between our laboratory findings and the RMF (NL3$^*$) model. We also observed a shell/sub-shell closure in terms of prolonged half-lives in isotopes with neutron numbers $N$ = 162, 164, and 174-184, which aligns with previous research. Furthermore, this study has shed light on the considerable differences in $\log_{10} T_{1/2}$ values when utilizing semi-empirical superheavy nuclei formulas. The comparison with previous research and the validation through standard deviation analysis have enhanced the confidence in the reliability of the MSLB results. Moreover, identifying isotopes with extended half-lives due to shell/sub-shell closures offers valuable information for future experiments in synthesizing superheavy nuclei. 

\vspace{-0.5cm}
\section*{Acknowledgments}
This work has been supported by the Science Engineering Research Board (SERB) File No. CRG/2021/001229, and FOSTECT Project No. FOSTECT.2019B.04.



\begin{thebibliography}{99}
\bibitem{sobi16}
A. Sobiczewski, ``Theoretical predictions for the nucleus $^{296}{118}$," Phys. Rev. C \textbf{94}(R), 051302 (2016).
\bibitem{manj19}
H. C. Manjunatha and K. N. Sridhar, ``A Detail Investigation on the Synthesis of Superheavy Element Z = 119," Phys. Part. Nucl. Lett. \textbf{16}, 647 (2019).
\bibitem{ogan17}
Y. T. Oganessian, A. Sobiczewski and G. M. Ter-Akopian, ``Superheavy nuclei: from predictions to discovery Phys. Scr.," \textbf{92}, 023003 (2017).
\bibitem{ogan09}
Y. T. Oganessian, et al., ``Attempt to produce element 120 in the $^{244}Pu$ + $^{58}Fe$ reaction," Phys. Rev. C \textbf{79}, 024603 (2009).
\bibitem{ogan12}
Y. T. Oganessian, et al., ``Production and Decay of the Heaviest Nuclei $^{293,294}117$ and $^{294}118$," Phys. Rev. Lett. \textbf{109}, 162501 (2012).
\bibitem{mose69}
U. Mosel and W. Greiner, ``On the stability of superheavy nuclei against fission," Z. Phys. \textbf{222}, 261 (1969).
\bibitem{bend01}
M. Bender et al., ``Shell stabilization of super-and hyperheavy nuclei without magic gaps," Phys. Lett. B \textbf{515}, 42 (2001).
\bibitem{cwio96}
S. Cwiok et al., ``Shell structure of the superheavy elements," Nucl. Phys. A \textbf{611}, 211 (1996).
\bibitem{seif11}
W. M. Seif, ``$\alpha$ decay as a probe of nuclear incompressibility," Phys. Rev. C \textbf{74}, 034302 (2006).
\bibitem{carr14}
R. J. Carroll, et al., ``Blurring the boundaries: Decays of multiparticle isomers at the proton drip line," Phys. Rev. Lett. \textbf{112}, 092501 (2014).
\bibitem{poen06}
D. N. Poenaru, I. H. Plonski, R. A. Gherghescu, and W. Greiner,  ``Valleys due to Pb and Sn on the potential energy surface of superheavy and lighter $\alpha$-emitting nuclei," J. Phys. G \textbf{32}, 1223 (2006).
\bibitem{park05}
A. Parkhomenko and A. Sobiczewski, ``Description of alpha spectroscopic data of odd-A superheavy nuclei," Acta Phys. Pol. B \textbf{36}, 1363 (2005).
\bibitem{mori07}
K. Morita, et al., ``Observation of second decay chain from $^{278}113$," J. Phys. Soc. Jpn. \textbf{76}, 045001 (2007).
\bibitem{ogan11}
Y. T. Oganessian, ``Synthesis of the heaviest elements in $^{48}$Ca-induced reactions," Radiochim. Acta \textbf{99}, 429 (2011).
\bibitem{dong09}
J. M. Dong, H. F. Zhang, and G. Royer, ``Proton radioactivity within a generalized liquid drop model," Phys. Rev. C \textbf{79}, 054330 (2009).
\bibitem{poen18}
D. N. Poenaru, H. St$\ddot{o}$cker, and R. A. Gherghescu, ``Cluster and alpha decay of superheavy nuclei," Eur. Phys. J. A \textbf{54}, 14 (2018).
\bibitem{deng19}
J. -G. Deng, X. -H. Li, J.-L. Chen, J. -H. Cheng, and X. -J. Wu, ``Systematic study of proton radioactivity of spherical proton emitters within various versions of proximity potential formalisms," Eur. Phys. J. A \textbf{55}, 58 (2019).
\bibitem{nave20}
G. Naveya, S. Santhosh Kumar, and A. Stephen, ``A systematic study on $\alpha$-decay chains of superheavy nuclei $Z = 126$ \& $138$," Int. Jour. Mod. Phys. E \textbf{29}, 2050034 (2020). 
\bibitem{hoff00}
D. C. Hoffman, A. Ghiorso, and G. T. Seaborg, ``The Transuranium People: The Inside Story" (Imperial College Press, London, 2000).
\bibitem{ogan07a}
Y. Oganessian, ``Heaviest nuclei from $^{48}$Ca-induced reactions," J. Phys. G \textbf{34}, R165 (2007).
\bibitem{khuy20}
 J. Khuyagbaatar, et al., ``Search for elements 119 and 120," Phys. Rev. C \textbf{102}, 064602 (2020).
\bibitem{xu06}
F. R. Xu, J. C. Pei, ``Mean-field cluster potentials for various cluster decays," Phys. Lett. B \textbf{642}, 322 (2006).
\bibitem{bhat08}
M. Bhattacharya, G. Gangopadhyay, ``$\alpha$-decay lifetime in superheavy nuclei with  A $>$ 282," Phys. Rev. C \textbf{77}, 047302 (2008).
\bibitem{dong10}
J. Dong, H. Zhang, Y. Wang, W. Zuo, J. Li, ``Alpha-decay for heavy nuclei in the ground and isomeric states," Nucl. Phys. A \textbf{832}, 198 (2010).
\bibitem{deli06}
D. S. Delion, S. Peltonen, J. Suhonen, ``Systematics of the $\alpha$-decay to rotational states," Phys. Rev. C \textbf{73}, 014315 (2006).
\bibitem{pelt07}
S. Peltonen, D. S. Delion, J. Suhonen, ``Folding description of the fine structure of $\alpha$ decay to 2$^+$ vibrational and transitional states," Phys. Rev. C \textbf{75}, 054301 (2007).
\bibitem{qian12}
Y. Qian and Z. Ren, ``Unfavored $\alpha$ decay from ground state to ground state in the range 53 $\leq$ Z $\leq$ 91," Phys. Rev. C {\bf 85}, 027306 (2012).
\bibitem{akra18}
D. T. Akrawy and A. H. Ahmed, ``New empirical formula for $\alpha$-decay calculations," Int. J. Mod. Phys. E {\bf 27}, 1850068 (2018).
\bibitem{seab90}
G. T. Seaborg and W. D. Loveland, ``The Elements Beyond Uranium" (Wiley-Interscience, New York, 1990).
\bibitem{myer66}
W. D. Myers and W. J. Swiatecki, ``Nuclear masses and deformations," Nucl. Phys. \textbf{81}, 1 (1966).
\bibitem{sobi66}
A. Sobiczewski, F. Gareev, and B. Kalinkin, ``Closed shells for Z $>$ 82 and N $>$ 126 in a diffuse potential well," Phys. Lett. \textbf{22}, 500 (1966).
\bibitem{viol66}
V. Viola and G. Seaborg, ``Nuclear systematics of the heavy elements—II Lifetimes for alpha, beta and spontaneous fission decay," J. Inorg. Nucl. Chem. \textbf{28}, 741 (1966).
\bibitem{gupt97}
R. K. Gupta, S. K. Patra, W. Greiner, ``Structure of $^{294,302}120$ nuclei using the relativistic mean-field method," Mod. Phys. Lett. A {\bf 12}, 1727 (1997).
\bibitem{patr99}
S. K. Patra, C. -L. Wu, C. R. Praharaj, R. K. Gupta, ``A systematic study of superheavy nuclei for Z = 114 and beyond using the relativistic mean field approach," Nucl. Phys. A {\bf 651}, 117 (1999).
\bibitem{ahma12}
S. Ahmad, M. Bhuyan, and S. K. Patra, ``Properties of Z = 120 nuclei and the $\alpha$-decay chains of the $^{292,304}120$ isotopes using Relativistic and Non-Relativistic formalism," Int. J. Mod. Phys. E {\bf 21},1250092 (2012).
\bibitem{tani20}
A. Taninah, S. E. Agbemava, and A. V. Afanasjev, ``Covariant density functional theory input for r-process simulations in actinides and superheavy nuclei: The ground state and fission properties," Phys. Rev. C {\bf 102}, 054330 (2020).
\bibitem{jerk18}
P. Jerabek, B. Schuetrumpf, P. Schwerdtfeger, and W. Nazarewicz, ``Electron and Nucleon Localization Functions of Oganesson: Approaching the Thomas-Fermi Limit," Phys. Rev. Lett. \textbf{120}, 053001 (2018).
\bibitem{dull18}
C. E. D$\ddot{u}$llmann and M. Block, ``Island of heavy weights," Sci. Am. \textbf{318}, 46 (2018).
\bibitem{bisw21}
N. Biswal, N. Jain, R. Kumar, A. S. Pradeep, S. Mishra, and M. Bhuyan, ``Structural and decay properties of nuclei appearing in the $\alpha$-decay chains of $^{296,298,300,302,304}$120 within the relativistic mean field formalism," Mod. Phys. Lett. A \textbf{36}, 2150169 (2021).
\bibitem{jain22}
N. Jain, R. Kumar, and M. Bhuyan, ``Exploring the ground state bulk and decay properties of the nuclei in superheavy island," Nucl. Phys. A, \textbf{1019},122379 (2022).
\bibitem{akra19}
D. Akrawy, and A. Ahmed, ``$\alpha$-decay systematics for superheavy nuclei," Phys. Rev. C \textbf{100}, 044618 (2019).
\bibitem{lala09}
G. A. Lalazissis, S. Karatzikos, R. Fossion, D. Pena Arteaga, A. V. Afanasjev, and P. Ring, ``The effective force NL3 revisited," Phys. Lett. B {\bf 671}, 36 (2009).
\bibitem{lala97}
G. A. Lalazissis, J. K$\ddot{o}$nig, and P. Ring, ``New parametrization for the Lagrangian density of relativistic mean field theory," Phys. Rev. C {\bf 55}, 540 (1997).
\bibitem{bogu77}
J. Boguta and A. R. Bodmer, ``Relativistic calculation of nuclear matter and the nuclear surface," Nucl. Phys. A {\bf 292}, 413 (1977).
\bibitem{sero86}
B. D. Serot and J. D. Walecka, ``In  Advances in Nuclear Physics, edited by J. W. Negele and Erich Vogt" (Plenum Press, New York, 1986), Vol {\bf 16}, p. 1.
\bibitem{bhuy11}
M. Bhuyan, S. K. Patra, and R. K. Gupta, ``Relativistic mean-field study of the properties of Z = 117 nuclei and the decay chains of the $^{293,294}117$ isotopes," Phys. Rev. C {\bf 84}, 014317 (2011).
\bibitem{bhuy15}
M. Bhuyan, ``Structural evolution in transitional nuclei of mass 82 $\leq$ A $\leq$ 132," Phys. Rev. C {\bf 92}, 034323 (2015).
\bibitem{type99}
S. Typel and H. H. Wolter, ``Relativistic mean field calculations with density-dependent meson-nucleon coupling," Nucl. Phys. A {\bf 656}, 331 (1999).
\bibitem{niks02}
T. Niksic, D. Vretenar, P. Finelli, and P. Ring, ``Relativistic Hartree-Bogoliubov model with density-dependent meson-nucleon couplings," Phys. Rev. C {\bf 66}, 024306 (2002).
\bibitem{ring96}
P. Ring, ``Relativistic mean field theory in finite nuclei," Prog. Part. Nucl. Phys. {\bf 37}, 193 (1996).
\bibitem{paar07}
N. Paar, D. Vretenar, and G. Colo, ``Exotic modes of excitation in atomic nuclei far from stability," Rep. Prog. Phys. {\bf 70}, 691 (2007).
\bibitem{bend03}
M. Bender, P. -H. Heenen, and P. -G. Reinhard, ``Self-consistent mean-field models for nuclear structure," Rev. Mod. Phys. {\bf 75},  121 (2003).
\bibitem{niks08}
T. Niksic, D. Vretenar, and P. Ring, ``Relativistic nuclear energy density functionals: Adjusting parameters to binding energies," Phys. Rev. C {\bf 78}, 034318 (2008).
\bibitem{fuch95}
C. Fuchs, H. Lenske, and H. H. Wolter, ``Density dependent hadron field theory," Phys. Rev. C {\bf 52}, 3043 (1995).
\bibitem{broc92}
R. Brockmann and H. Toki, ``Relativistic density-dependent Hartree approach for finite nuclei," Phys. Rev. Lett. {\bf 68}, 3408 (1992).
\bibitem{niko92}
B. A. Nikolaus, T. Hoch, and D. G. Madland, ``Nuclear ground state properties in a relativistic point coupling model," Phys. Rev. C {\bf 46},1757 (1992).
\bibitem{burv02}
T. Burvenich, D. G. Madland, J. A. Maruhn, and P. -G. Reinhard, ``Nuclear ground state observables and QCD scaling in a refined relativistic point coupling model," Phys. Rev. C {\bf 65}, 044308 (2002).
\bibitem{rein89}
P. -G. Reinhard, ``The relativistic mean-field description of nuclei and nuclear dynamics," Rep. Prog. Phys. {\bf 52}, 439 (1989).
\bibitem{vret05}
D. Vretenar, A. V. Afanasjev, G. A. Lalazissis, and P. Ring, ``Relativistic Hartree–Bogoliubov theory: static and dynamic aspects of exotic nuclear structure," Phys. Rep. {\bf 409}, 101 (2005).
\bibitem{bhuy18}
M. Bhuyan, B. V. Carlson, S. K. Patra, and S. -G. Zhou, ``Surface properties of neutron-rich exotic nuclei within relativistic mean field formalisms," Phys. Rev. C {\bf 97}, 024322 (2018).
\bibitem{patr09}
S. K. Patra, M. Bhuyan, M. S. Mehta, and R. K. Gupta, ``Superdeformed and hyperdeformed states in Z = 122 isotopes," Phys. Rev. C {\bf 80}, 034312 (2009).
\bibitem{niks11}
T. Niksic, D. Vretenar, and P. Ring, ``Relativistic nuclear energy density functionals: Mean-field and beyond," Prog. Part. Nucl. Phys. {\bf 66}, 519 (2011).
\bibitem{long04}
W. Long, J. Meng, N. Van Giai, and S. -G Zhou, ``New effective interactions in relativistic mean field theory with nonlinear terms and density-dependent meson-nucleon coupling," Phys. Rev. C {\bf 69},  034319 (2004).
\bibitem{pann86}
W. Pannert, P. Ring, and J. Boguta, ``Relativistic mean-field theory and nuclear deformation," Phys. Rev. Lett. {\bf 59}, 2420 (1986).
\bibitem{kara10}
S. Karatzikos, A. V. Afanasjev, G. A. Lalazissis, and P. Ring, ``The fission barriers in Actinides and superheavy nuclei in covariant density functional theory," Phys. Lett. B {\bf 689}, 72 (2010).
\bibitem{lunn03}
D. Lunney, J. M. Pearson and C. Thibault, ``Recent trends in the determination of nuclear masses," Rev. Mod. Phys. {\bf 75}, 1021 (2003).
\bibitem{lala99}
G. A. Lalazissis, S. Raman, and P. Ring, ``Ground-state properties of even–even nuclei in the relativistic mean-field theory," At. Data Nucl. Data Tables {\bf 71}, 1 (1999).
\bibitem{zhan06}
H. F. Zhang, J. Q. Li, W. Zuo, X. H. Zhou, Z. G. Gan, and N. Wang, ``Ground state properties of superheavy nuclei in relativistic mean field theory," Int.  J. Mod.  Phys. {\bf  15}, 1613 (2006).
\bibitem{sahu11}
B. K. Sahu, M. Bhuyan, S. Mahapatro, and S. K. Patra, ``The $\alpha$- decay chains of the 287,288115 isotopes using relativistic mean field theory," Int. J. Mod.  Phys.  E {\bf 20}, 2217 (2011).
\bibitem{bhuy15a}
M. Bhuyan, S. Mahapatro, and S. Singh, S. K. Patra, ``The structural and decay properties of Francium isotopes," Int. J. Mod. Phys. E \textbf{24}, 1550028 (2015).
\bibitem{bhuy18a}
M. Bhuyan, ``Probable Decay Modes at Limits of Nuclear Stability of the Superheavy Nuclei," Phys. Atom. Nuclei \textbf{81}, 15 (2018).
\bibitem{moll19}
P. M$\ddot{o}$ller, M. Mumpower, T. Kawano, W. Myers, ``Nuclear properties for astrophysical and radioactive-ion-beam applications (II)," At. Data Nucl. Data Tables \textbf{125}, 1 (2019).
\bibitem{wang14}
N. Wang, M. Liu, X. Wu and J. Meng, ``Surface diffuseness correction in global mass formula," At. Data Nucl. Data Tables Phys. Lett. B {\bf 734}, 215 (2014).
\bibitem{wang12}
M. Wang, G. Audi, A. H. Wapstra, F. G. Kondev, M. MacCormick, X. Xu and B. Pfeiffer, ``The Ame2012 atomic mass evaluation," Chin. Phys. C {\bf 36}, 1603 (2012).
\bibitem{frdm97} 
P. M\~oller, J. R. Nix, and K. -L. Kratz, ``Nuclear properties for astrophysical and radioactive-ion-beam applications.," {\it At. Data and Nucl. Data Tables} {\bf 66}, 131 (1997).
\bibitem{frdm16} 
P. M\~oller, A. J. Sierk, T. Ichikawa {\it et al.}, ``Nuclear ground-state masses and deformations: FRDM (2012).," {\it At. Data and Nucl. Data Tables} {\bf 109}, 1 (2016).
\bibitem{rama01}
S. Raman, C. W. Nestor, JR., and P. Tikkanen, ``Transition probability from the ground to the first-excited $2^+$ state of even-even nuclides"$*$", At. Data Nucl. Data Tables  {\bf 78}, 1 (2001).
\bibitem{lu16}
Bing-Nan Lu, Jie Zhao, En-Guang Zhao, and Shan-Gui Zhou, ``Relativistic Density Functional for Nuclear Structure", Ch. 05, pp. 171-217 (2016).
\bibitem{geig11}
H. Geiger and J. M. Nuttall, ``LVII. The ranges of the $\alpha$ particles from various radioactive substances and a relation between range and period of transformation," Phil. Mag. {\bf 22}, 613 (1911).
\bibitem{Roye00}
G. Royer, ``Alpha emission and spontaneous fission through quasi-molecular shapes," J. Phys. G: Nucl. Part. Phys. \textbf{26}, 1149 (2000). 
\bibitem{Roye10}
G. Royer, ``Analytic expressions for alpha-decay half-lives and potential barriers," Nucl. Phys. A \textbf{848}, 279 (2010). 
\bibitem{Poen07}
D. Poenaru, R. Gherghescu, N. and Carjan, ``Alpha-decay lifetimes semi-empirical relationship including shell effects," Euro. phys. Lett. \textbf{77}, 62001 (2007).
\bibitem{Manj17}
H. Manjunatha, K. Sridhar, ``New semi-empirical formula for
$\alpha$-decay half-lives of the heavy and superheavy nuclei," Eur. Phys. J. A \textbf{53}, 1 (2017).
\bibitem{Akra17}
D. Akrawy, D. Poenaru, ``Alpha decay calculations with a new formula," J. Phys. G: Nucl. Part. Phys. \textbf{44}, 105105 (2017).
\bibitem{Akra18b}
D. Akrawy, A. Ahmed, ``New empirical formula for $\alpha$-decay calculations," Int. J. Mod. Phys. E \textbf{27}, 1850068 (2018).
\bibitem{hofm15}
S. Hofmann, ``Super-heavy nuclei." J. Phys. G: Nucl. Part. Phys. {\bf {42}}, 114001 (2015).
\bibitem{giul19}
S. A. Giuliani, Z. Matheson, W. Nazarewicz, E. Olsen, P. G. Reinhard, J. Sadhukhan, B. Schuetrumpf, N. Schunck, N. and P. Schwerdtfeger, ``Colloquium: Superheavy elements: Oganesson and beyond" Rev. Mod. Phys. \textbf{91}, 011001 (2019).
\bibitem{thee23}
Theeb Y. T. Alsultan, T. M. Joshua, R. Kumar, B. T. Goh, and M. Bhuyan, ``Impact of nuclear rotation corrections on alpha decay half-lives of superheavy nuclei within 98 $\leq$ Z $\leq$ 120" Nucl. Phys. A, {\bf 1041}, 122784 (2023). 
\bibitem{audi17}
G. Audi, F. G. Kondev, M. Wang, W. J. Huang, and S. Naimi, ``The NUBASE2016 evaluation of nuclear properties," At. Data Nucl. Data Tables Chin. Phys. C {\bf 41}, 030001 (2017).
\end{thebibliography}
\end{document}